\documentclass[aps,prd,twocolumn,nofootinbib,showpacs]{revtex4}
\usepackage[dvips]{graphicx}
\usepackage{amssymb}
\usepackage{psfrag}
\usepackage{array}
\usepackage{float}
\usepackage{epsfig}
\usepackage{amssymb,amsmath}
\usepackage{hyperref}
\usepackage{multirow}
\allowdisplaybreaks[2]

\newcommand{\nn}{\nonumber} 
\newcommand{\bea}{\begin{eqnarray}}
\newcommand{\eea}{\end{eqnarray}}
\newcommand{\beq}{\begin{equation}}
\newcommand{\eeq}{\end{equation}}
\newcommand{\phie}{\sigma}

\begin{document}

\title{Non-Minimal $B-L$ Inflation with Observable Gravity Waves}

\author{Nobuchika Okada$^a$} \email{E-mail: okadan@ua.edu}
\author{Mansoor Ur Rehman$^b$} \email{E-Mail: rehman@udel.edu} 
\author{Qaisar Shafi$^b$} \email{E-mail: shafi@bartol.udel.edu}
\affiliation{$^a$Department of Physics and Astronomy, University of Alabama, 
Tuscaloosa, AL 35487, USA}
\affiliation{$^b$Bartol Research Institute, Department of Physics and Astronomy,
University of Delaware, Newark, DE 19716, USA}

%%%%%%%%%%%%%%%%%%%%%%%%
\begin{abstract}
%%%%%%%%%%%%%%%%%%%%%%%%

We consider non-minimal $\lambda \phi^4$ inflation in a 
gauged non-supersymmetric $U(1)_{B-L}$ model containing the 
gravitational coupling $\xi\,\mathcal{R}\,\Phi^\dagger \Phi $, where 
$\mathcal{R}$ denotes the Ricci scalar and the standard model 
singlet inflaton field $\Phi$ spontaneously breaks the 
$U(1)_{B-L}$ symmetry. Including radiative corrections, 
the predictions $0.956 \lesssim n_s \lesssim 0.984$ and 
$0.007 \lesssim r \lesssim 0.1$ for the scalar spectral 
index and tensor to scalar ratio $r$ lie within the current 
WMAP 1-$\sigma$ bounds. 
If the $B-L$ symmetry breaking scale is of order a TeV or so, 
one of the three right handed neutrinos is a plausible 
cold dark matter candidate. Bounds on the dimensionless
parameters $\lambda$, $\xi$ and the gauge coupling 
$g_{B-L}$ are obtained.

%%%%%%%%%%%%%%%%%%%%%%%%
\end{abstract}
%%%%%%%%%%%%%%%%%%%%%%%%
\maketitle

A non-minimal gravitational coupling of inflaton
has been known to play an important role in models
of chaotic inflation \cite{Salopek:1988qh}. 
Recently, this idea has received a fair amount of attention
\cite{Bezrukov:2008dt}-\cite{Linde:2011nh}
arising from the possibility of taking the Standard Model (SM) Higgs 
boson as an inflaton.
In the simplest scenarios of this kind, the SM Higgs doublet 
$H$ has a relatively strong non-minimal gravitational interaction 
$\xi\,\mathcal{R}\,H^\dagger H $, where $\mathcal{R}$ is the Ricci 
scalar and $\xi$ a dimensionless coupling whose magnitude 
is estimated to be of order $10^4$ 
based on WMAP data \cite{Komatsu:2010fb}. 
This SM Higgs-based inflationary scenario is currently mired 
in some controversy stemming from the arguments put forward in 
\cite{cutoff} that for $\xi \gg 1$, the energy scale 
$\lambda^{1/4}\,m_P / \sqrt{\xi}$ during non-minimal SM inflation 
exceeds the effective ultraviolet cut-off scale  
$\Lambda = m_P/\xi$. 
Here $\lambda$ of order unity denotes the SM Higgs 
quartic coupling and 
$m_P \simeq 2.43 \times 10^{18}$~GeV represents 
the reduced Planck mass. This point has been further 
elaborated in Refs.~\cite{Burgess:2010zq,Hertzberg:2010dc}.
However, it has recently been argued \cite{Bezrukov:2010jz} 
that if the Higgs field is perturbed around 
its non-zero classical background, the effective cut-off
can become larger than the energy scale of inflation.
(See \cite{Germani:2010gm} for other possible solutions
to the unitarity problem.)
As we will see below, the above problem has a negligible 
impact on our conclusions in this paper. Even though
the inflation in our case carries a $U(1)_{B-L}$ charge,
a satisfactory scenario imposes relatively mild but
nonetheless important constraints on the $U(1)_{B-L}$
gauge coupling.

In this paper we implement non-minimal $\phi^4$ inflation by 
supplementing the standard model with a gauged $U(1)_{B-L}$ 
symmetry \cite{Pati:1974yy}. (For inflation with 
local supersymmetric $U(1)_{B-L}$, see 
\cite{Lazarides:1996dv,Jeannerot:1997av,Senoguz:2005bc} 
and references therein. For global $U(1)_{B-L}$ inflation 
see \cite{Shafi:1984tt}) 
The well-known advantages of a spontaneously broken gauge 
$U(1)_{B-L}$ symmetry include seesaw physics \cite{Seesaw} 
to explain neutrino oscillations, and baryogenesis via 
leptogenesis \cite{Fukugita:1986hr,Lazarides:1991wu}
arising from the right handed neutrinos that are 
present to cancel the gauge anomalies.
In the inflation model that we consider the symmetry 
breaking scale of $U(1)_{B-L}$ is arbitrary as long as
the lower bound from LEP experiment, $\gtrsim 3$ TeV 
\cite{LEPbound}, is satisfied.
One interesting possibility is to break it at the
TeV scale \cite{Khalil:2006yi}, and it has been 
shown \cite{Iso:2009ss,Iso:2009nw} 
that the minimal $U(1)_{B-L}$ model with additional 
classical conformal invariance naturally predicts the 
symmetry breaking scale to be at TeV. 
This means that the new particles, the $Z'$ gauge boson, 
the $B-L$ Higgs boson $\Phi$ and the RH neutrinos $N^i$ 
have TeV scale masses, and they can be observed at 
Large Hadron Collider 
(LHC)~\cite{Emam:2007dy,Huitu:2008gf,Basso:2008iv,Perez:2009mu}.
Furthermore, with TeV scale RH neutrinos we can
explain the origin of the baryon asymmetry through
resonant leptogenesis \cite{resonantLG, resonantLG2}.

One important feature missing in the above 
TeV scale $U(1)_{B-L}$ model is non-baryonic
dark matter (DM). 
To circumvent this, following Ref.~\cite{Okada:2010wd}, 
we introduce an unbroken $Z_2$ parity under which
one of the three RH neutrinos is taken to be 
odd, while all other fields are even. In this case the 
$Z_2$-odd RH neutrino is absolutely stable 
and a viable DM candidate. 
Note that the two remaining RH neutrinos are sufficient to 
reconcile theory with the observed neutrino oscillation data. 
The model also predicts that one of the three observed neutrinos
is essentially massless.
Thus, without introducing any additional 
dynamical degrees of freedom, the DM particle can be
incorporated in the minimal gauged $U(1)_{B-L}$ model.

In this paper we consider non-minimal $\lambda\,\phi^4$ inflation 
by taking $\phi$ ($ \equiv  \sqrt{2}\,Re(\Phi)$), %\phi\,e^{i\theta}/\sqrt{2}$) 
to be the inflaton field which is charged under $B-L$. 
We take into account quantum corrections 
to the inflationary potential arising from the inflaton interactions 
with the $U(1)_{B-L}$ gauge field. We find that the
tensor to scalar ratio $r \gtrsim 0.007$ 
and the scalar spectral index $n_s \gtrsim 0.956$.
More generally, in this non-minimal $\lambda \phi^4$ inflation model, 
the predictions $0.956 \lesssim n_s \lesssim 0.984$ and 
$0.007 \lesssim r \lesssim 0.1$ lie within the WMAP 1-$\sigma$ 
bounds for $10^{-12} \lesssim \lambda \lesssim 0.3$ and 
$10^{-3} \lesssim \xi \lesssim 10^4$. 
Recall that the corresponding tree level predictions for minimal 
($\xi = 0$) $\lambda\,\phi^4$ chaotic inflation, namely $n_s \simeq 0.95$ 
and $r \simeq 0.26$, lie outside the WMAP 2-$\sigma$ bounds.

Our inflation model is based on the gauge group 
$SU(3)_c \times SU(2)_L \times U(1)_Y \times U(1)_{B-L}$ 
and the particle content is listed in Table~1 
\cite{Iso:2009nw}.
The SM singlet scalar ($\Phi$) breaks the $U(1)_{B-L}$ 
gauge symmetry down to $Z_{2\,(B-L)}$ by its vacuum expectation 
value (vev), and at the same time generates the 
right-handed neutrino masses. The Lagrangian terms relevant for the 
seesaw mechanism are given by 
\bea 
 {\cal L} \supset -Y_D^{ij} \overline{N^i} H^\dagger \ell_L^j  
- \frac{1}{2} Y_N^i \Phi \overline{N^{i^{c}}} N^i 
+{\rm h.c.},  
\label{Yukawa}
\eea
where the first term yields the Dirac neutrino mass 
after electroweak symmetry breaking, 
while the right-handed neutrino Majorana mass term 
is generated by the second term associated with 
the $B-L$ gauge symmetry breaking. 
Without loss of generality, we use the basis
which diagonalizes the second term and makes
$Y_N^i$ real and positive.

%%%%%%%%%%%%%%%%%%%%%%%%%%%%%%%%%%%%%%%%%%%%%%%
\begin{table}[t]
\begin{center}
\begin{tabular}{c|ccc|c}
            & SU(3)$_c$ & SU(2)$_L$ & U(1)$_Y$ & U(1)$_{B-L}$  \\
\hline
$ q_L^i $    & {\bf 3}   & {\bf 2}& $+1/6$ & $+1/3$  \\ 
$ u_R^i $    & {\bf 3} & {\bf 1}& $+2/3$ & $+1/3$  \\ 
$ d_R^i $    & {\bf 3} & {\bf 1}& $-1/3$ & $+1/3$  \\ 
\hline
$ \ell^i_L$    & {\bf 1} & {\bf 2}& $-1/2$ & $-1$  \\ 
$ N^i$   & {\bf 1} & {\bf 1}& $ 0$   & $-1$  \\ 
$ e_R^i  $   & {\bf 1} & {\bf 1}& $-1$   & $-1$  \\ 
\hline 
$ H$         & {\bf 1} & {\bf 2}& $-1/2$  &  $ 0$  \\ 
$ \Phi$      & {\bf 1} & {\bf 1}& $  0$  &  $+2$  \\ 
\end{tabular}
\end{center}
\caption{
Particle content. 
In addition to the SM particle contents, 
the right-handed neutrino $N^i$ 
($i=1,2,3$ denotes the generation index) 
and a complex scalar $\Phi$ are introduced. 
}
\end{table}
%%%%%%%%%%%%%%%%%%%%%%%%%%%%%%%%%%%%%%%%%%%%%%%

Consider the following tree level action in the Jordan frame:
\bea  \label{action1}
S_J^{tree} && = \int d^4 x \sqrt{-g} 
\left[- \left( \frac{m_P^2 }{2}+ \xi_{H} H^{\dagger} H
+ \xi \Phi^{\dagger} \Phi \right)\mathcal{R} \right. \nn \\ 
&&  +(D_{\mu} H)^{\dagger}g^{\mu \nu}(D_{\nu} H) 
- \lambda_H \left( H^{\dagger} H - \frac{v^2}{2} \right)^2 \nn \\ 
&& + (D_{\mu} \Phi)^{\dagger}g^{\mu \nu}(D_{\nu} \Phi) 
- \lambda \left( \Phi^{\dagger} \Phi - \frac{v_{B-L}^2}{2} \right)^2 \nn \\ 
&& \left. -\lambda' (\Phi^{\dagger}\Phi)(H^{\dagger}H)  \right],
\eea
where $v$ and $v_{B-L}$ are the vevs of the Higgs fields 
$H$ and $\Phi$ respectively. 
To simplify the discussion, we assume that 
$\lambda'$ is sufficiently small so it can be ignored,
and also $\xi_H \ll \xi$.

The relevant one-loop renormalization group improved effective 
action can be written as \cite{RGEIP}
\bea
S_J &=& \int d^4 x \sqrt{-g} 
\left[- \left( \frac{m_P^2 + 
\xi\,G(t)^2\phi^2}{2}\right)\mathcal{R} \right. \nn \\
&& \left. + \frac{1}{2}G(t)^2(\partial \phi)^2 
- \frac{1}{4} \lambda(t) G(t)^4 \phi^4  \right],
\eea
where $t=\ln(\phi/\mu)$ and $G(t)= \exp(- \, \int_0^t dt'
\gamma(t')/(1+\gamma(t')))$, with 
\beq
\gamma(t) = \frac{1}{(4\pi)^2} 
\left(\frac{1}{2}\sum_i (Y_N^i(t))^2 - 12\,g_{B-L}^2(t) \right)
\eeq
being the anomalous dimension of the inflaton field.
$g_{B-L}$ denotes the $U(1)_{B-L}$ gauge coupling and $\mu$
the renormalization scale.
In the presence of the nonminimal gravitational coupling, 
the one loop renormalization group equations (RGEs)
of $\lambda$, $g_{B-L}$, $\xi$ and $Y^i_N$ are 
given by \cite{Iso:2009ss,Iso:2009nw}
\bea
(4\pi)^2\,\frac{d\lambda}{dt} &=& (2+18\,s^2)\,\lambda^2 
- 48 \lambda\,g^2_{B-L} +  96 g_{B-L}^4
 \nn \\
&& + 2 \lambda \,\sum_i (Y_N^i)^2 - \sum_i (Y_N^i)^4, \\
(4\pi)^2\,\frac{dg_{B-L}}{dt} &=& 
\left( \frac{32 + 4\,s}{3}\right)\,g^3_{B-L},  \\
(4\pi)^2\,\frac{d\xi}{dt} &=& 
(\xi + 1/6)\left( (1+s^2)\lambda -2 \gamma \right),  \\
(4\pi)^2\,\frac{dY_N^i}{dt} &=& (Y_N^i)^3
 - 6 g^2_{B-L} Y_N^i 
 +\frac{1}{2} Y_N^i  \sum_j (Y_N^j)^2, \nn  \\
 &&  %+\frac{1}{2} Y_N^i  \sum_j (Y_N^j(t))^2
\eea
where the $s$ factor is defined as
\beq
s(\phi) \equiv
 \frac{\left( 1 + \frac{\xi \phi^2}{m_P^2} \right)}{1+(6 \xi +1)\frac{\xi
 \phi^2}{m_P^2}}.  \label{s}
\eeq

In the Einstein frame with a canonical gravity sector, 
the kinetic energy of $\phi$ can be made canonical with respect 
to a new field $\phie = \phie (\phi)$ \cite{Clark:2009dc}, 
\beq
\left({d\phie\over d\phi}\right)^2 =
\frac{G(t)^2\Omega(t)+3m_P^2(\partial_{\phi}\Omega(t))^2/2}{\Omega(t)^2},  \label{kinetic}
\eeq
where,
\beq
\Omega(t) = 1 + \xi\,G(t)^2\phi^2/m_P^2.
\eeq
The action in the Einstein frame is then given by
\beq
S_E = \!\int d^4 x \sqrt{-g_E}\left[-\frac12 m_P^2 \mathcal{R}_E+\frac12 (\partial_E \phie)^2
-V_E(\phie)\right],
\label{action}
\eeq
with
\beq
V_E(\phi) = \frac{\frac{1}{4} \lambda(t)\,G(t)^4\,\phi^4}{\left(1+\frac{\xi\,\phi^2}{m_P^2}\right)^2}.
\label{potrgi}
\eeq
In our numerical work, we employ above potential 
with the RGEs given in Eqs.~(5-8).
However, for a qualitative discussion it is reasonable to use 
the following leading-log approximation of the 
above potential: 
\beq
V_E(\phi) \simeq \frac{\left( \frac{\lambda_0}{4}
+ \frac{96\,g^2_{B-L}}{16\,\pi^2} 
\ln \left[ \frac{\phi}{\mu} \right] \right)\phi^4}
{\left(1+\frac{\xi\,\phi^2}{m_P^2}\right)^2}, 
\label{ApproxPotential}
\eeq
where we have assumed $\gamma \approx 0$,
$dY^i_N/dt \approx 0$, $dg_{B-L}/dt \approx 0$,
$d\xi/dt \approx 0$, $g^2_{B-L} \ll (\lambda,\,(Y_N^i)^2)$, 
and $d \lambda /dt \approx 96 g_{B-L}^4 /(4\pi)^2$
with $\lambda_0 \equiv \lambda(t=0)$. We have
checked that for a broad range of parameters
the above expression can be regarded as a valid 
approximation for the potential given in 
Eq.~(\ref{potrgi}). 
In our numerical calculations we fix the 
renormalization scale $\mu = 1$ TeV.

To discuss the predictions of this model it is useful to 
first recall the basic results of the slow roll assumption. 
The inflationary slow-roll parameters are given by
\bea
\epsilon(\phi)&=&\frac12 m_P^2 \left({V_E'\over V_E \sigma'}\right)^2 
,\label{epsilon}\\
\eta(\phi)&=&m_P^2\left[
{V_E''\over V_E (\sigma')^2}
\!-{V_E'\sigma''\over V_E (\sigma')^3}\right],\,\,\,\,\,\label{eta}  \\
\zeta^2 (\phi) &=& m_P^4 \left({V_E'\over V_E \sigma'}\right) \left( \frac{V_E'''}{V_E (\sigma')^3}
-3 \frac{V_E'' \sigma''}{V_E (\sigma')^4} \right. \\
&& \left. + 3 \frac{V_E' (\sigma'')^2}{V_E (\sigma')^5}
- \frac{V_E' \sigma'''}{V_E (\sigma')^4} \right)  , \label{xi}
\eea
where a prime denotes a derivative with respect to $\phi$. 
The slow-roll approximation is valid as long as the conditions 
$\epsilon \ll 1$, $|\eta| \ll 1$ and $\zeta^2 \ll 1$ hold. In this case 
the scalar spectral index $n_{s}$, the tensor-to-scalar ratio $r$, and the 
running of the spectral index $\frac{d n_{s}}{d \ln k}$ are 
approximately given by
\bea
n_s &\simeq& 1-6\,\epsilon + 2\,\eta, \\
r &\simeq& 16\,\epsilon,  \\ 
\frac{d n_{s}}{d \ln k} \!\!&\simeq&\!\!
16\,\epsilon\,\eta - 24\,\epsilon^2 - 2\,\zeta^2. \label{reqn}
\eea
The number of e-folds after the comoving scale $l$ has 
crossed the horizon is given by
\beq
N_l={1\over \sqrt{2}\, m_P}\int_{\phi_{\rm e}}^{\phi_l}
{d\phi\over\sqrt{\epsilon(\phi)}}\left({d\phie\over d\phi}\right)\,,
\label{Ne}\eeq
where $\phi_l$ is the field value at the comoving scale 
$l$, and $\phi_e$ denotes the value of $\phi$ at
the end of inflation, defined by 
max$(\epsilon(\phi_e) , |\eta(\phi_e)|,\zeta^2(\phi_e)) = 1$.

The amplitude of the curvature perturbation 
$\Delta_{\mathcal{R}}$ is given by 
\begin{equation} \label{Delta}
\Delta_{\mathcal{R}}^2 = \left. \frac{V_E}{24\,\pi^2 \, m_P^2\,\epsilon } \right|_{k_0},
\end{equation}
where $\Delta_{\mathcal{R}}^2 = (2.43\pm 0.11)\times 10^{-9}$ is 
the WMAP7 normalization at $k_0 = 0.002\, \rm{Mpc}^{-1}$ 
\cite{Komatsu:2010fb}. 
Note that for added precision, we include in our 
calculations the first order corrections \cite{Stewart:1993bc} 
in the slow-roll expansion for the quantities 
$n_s$, $r$, $\frac{dn_s}{d \ln k}$, and $\Delta_{\mathcal{R}}$.

Using Eqs.~(\ref{ApproxPotential})-(\ref{Delta}) above 
we can obtain various predictions of the radiatively 
corrected non-minimal $\phi^4$ model of inflation. 
Once we fix the parameters $\xi$ and $g_{B-L}$, 
and the number of e-foldings $N_0$, we can predict $n_s$, $r$, 
and $\frac{d n_{s}}{d \ln k}$. 
The tree level ($g_{B-L} = 0$) predictions  for minimal $\phi^4$ inflation
are readily obtained as:
\bea
n_s &=& 1 - \frac{24}{\phi^2} = 1 - \frac{3}{N_0}, 
\label{nsxi0} \\
r &=& \frac{128}{\phi^2} = \frac{16}{N_0}, 
\label{rxi0} \\
\frac{d n_{s}}{d \ln k}  &=& -\frac{192}{\phi^4} = -\frac{3}{N_0^2}.
\label{axi0}
\eea
For $N_0 = 60$ ($N_0 = 50$), we find 
$n_s \simeq 0.95$ ($n_s \simeq 0.94$), 
$r \simeq 0.26$ ($r \simeq 0.31$) and 
$\frac{d n_{s}}{d \ln k} \simeq -8 \times 10^{-3}$
($\frac{d n_{s}}{d \ln k} \simeq - 10^{-3}$). 
\begin{figure}[t]
\centering \includegraphics[width=8.25cm]{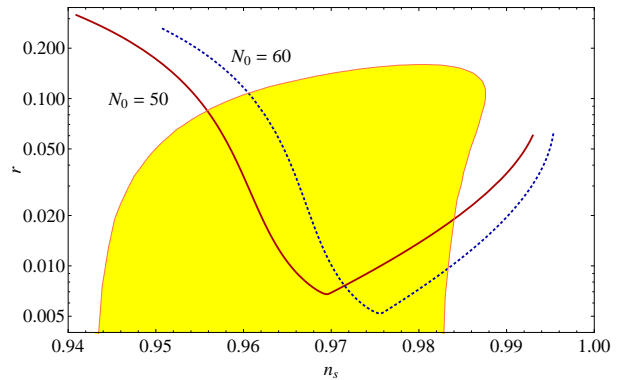}
\caption{$r$ vs. $n_s$ for the radiatively corrected 
non-minimal $\phi^4$ potential defined in Eq.~(\ref{ApproxPotential})
with the number of e-foldings $N_0 = 50$ (red solid curve) 
and $N_0 = 60$ (blue dashed curve) and $\lambda_0$ = 0. 
The WMAP 1-$\sigma$ (68\% confidence level) bounds are shown in yellow.
Along each curve we vary either $g_{B-L}$ or $\xi$.} \label{fig1}
\end{figure}
As expected, the predictions of tree 
level minimal $\phi^4$ inflation lie outside the 
2-$\sigma$ WMAP bounds \cite{Komatsu:2010fb}.

However, the situation is improved once the radiative 
corrections are included \cite{NeferSenoguz:2008nn}.
The impact of these radiative corrections on
the tree level predictions of various inflationary models
have been studied in Refs.~\cite{Rehman:2009wv,Rehman:2010es}.
Furthermore, the nonminimal gravitational coupling
also plays an important role in making the tree level 
predictions consistent with the WMAP data. 
Indeed, the radiative corrections smear out the
tree level predictions of nonminimal inflationary 
models \cite{Okada:2010jf}. A similar behavior is observed
in our situation. 
The approximate potential in Eq.~(\ref{ApproxPotential}) 
effectively behaves as a 
nonminimal $\lambda_{\phi}\phi^4/4$ potential with a running
coupling constant $\lambda_{\phi} \approx \lambda_0
+ \frac{96\,g^2_{B-L}}{4\,\pi^2}\ln \left[ \frac{\phi}{\mu} \right]$.
In the limit $\xi \ll 1$, assuming $\lambda_{\phi}$
to be approximately constant, the scalar spectral index, 
the tensor to scalar ratio and the 
running of the spectral index for the 
radiatively corrected non-minimal $\phi^4$ 
inflation are given by \cite{Okada:2010jf}
\bea
n_s &\simeq&  1 - \frac{3(1+16\,\xi N_0/3)}{N_0\,(1+8\,\xi N_0)}, 
\label{nssmallxi} \\
r &\simeq&  \frac{16}{N_0\,(1+8\,\xi N_0)}, \label{rsmallxi} \\
\frac{d n_{s}}{d \ln k}  &\simeq&  
\frac{r}{2} \left(\frac{16\,r}{3} -(1 - n_s)\right) \nn \\
&& \hspace{-2.0cm}
-\frac{3\,\left(1+4\,(8\,\xi N_0)/3-5\,(8\,\xi N_0)^2
-2\,(8\,\xi N_0)^3 \right)}{N_0^2\,(1+8\,\xi N_0)^4}.
\label{amallxi}
\eea
\begin{figure}[t]
\centering \includegraphics[width=8.25cm]{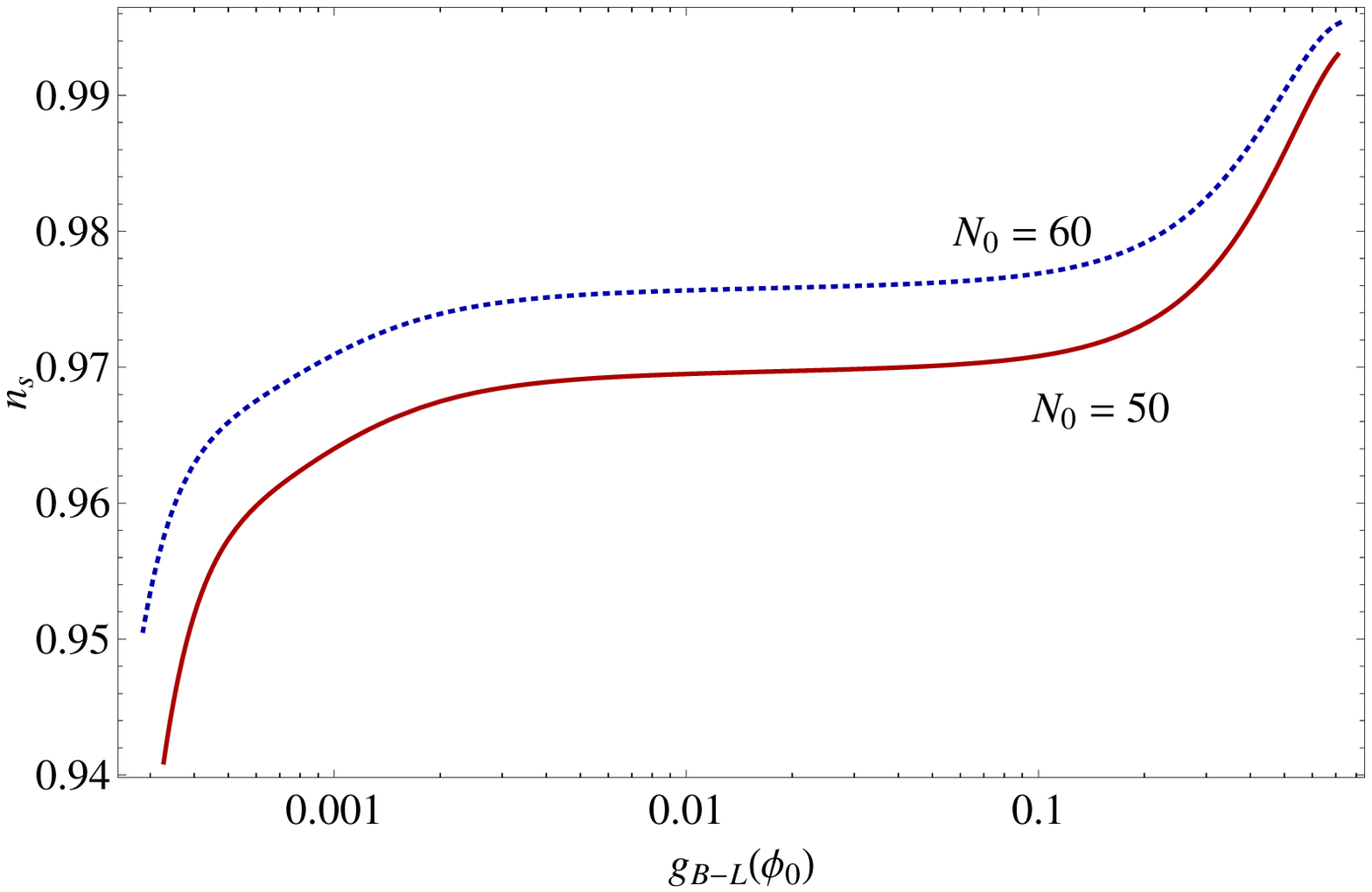}
\centering \includegraphics[width=8.25cm]{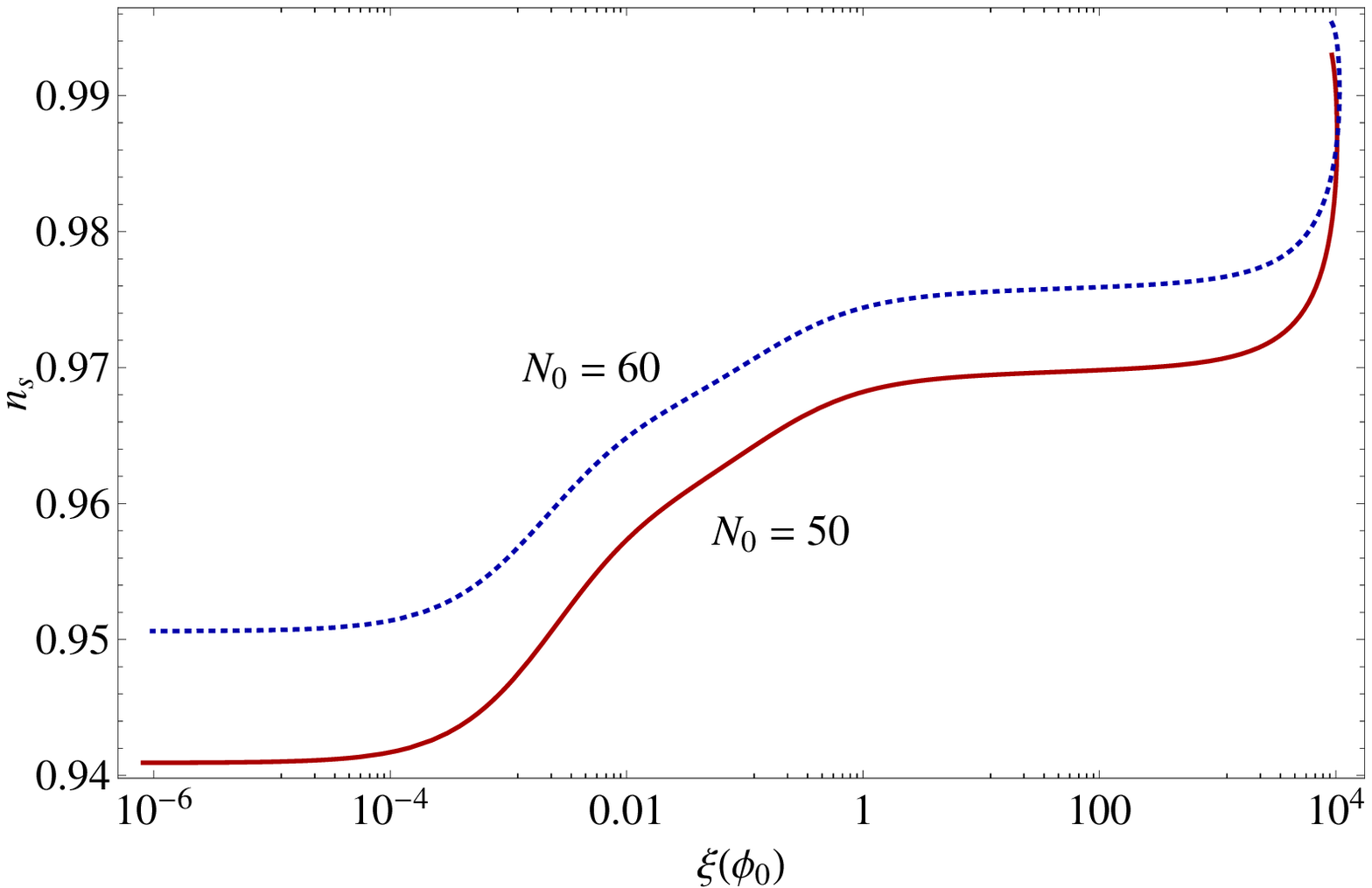}
\caption{$n_s$ vs. $g_{B-L}(\phi_0)$ and $\xi(\phi_0)$ for  
radiatively corrected non-minimal $\phi^4$ inflation with 
the number of e-foldings $N_0 = 50$ (red solid curve) 
and $N_0 = 60$ (blue dashed curve) 
and $\lambda_0$ = 0.} 
\label{fig2}
\end{figure}
These results exhibit a reduction in the value 
of $r$ and an increase in the value of $n_s$ 
compared to their minimally coupled tree level predictions 
(Eqs.~(\ref{nsxi0}-\ref{axi0})),
as can be seen in Figs.~\ref{fig1}-\ref{fig3}. 
In our analysis, we set $\lambda_0=0$ limit for simplicity. 
From the WMAP 1-$\sigma$ bounds 
($r \sim 0.1$ and $n_s \sim 0.96$), we obtain a 
lower bound of $\xi \gtrsim 3 \times 10^{-3}$ with 
$N_0 = 60$ e-foldings \cite{Bezrukov:2008dt}.
The value of $\frac{d n_{s}}{d \ln k}$ 
receives a tiny correction to its tree level prediction.
Note the sharp transitions in the predictions of $n_s$
and $r$ in the vicinity of $\xi \approx 10^{-2}$. This
can be understood from the expression for the inflationary
potential given in Eq.~(\ref{ApproxPotential}) 
and Eqs.~(\ref{nssmallxi}) and (\ref{rsmallxi}).

\begin{figure}[t]
\centering \includegraphics[width=8.25cm]{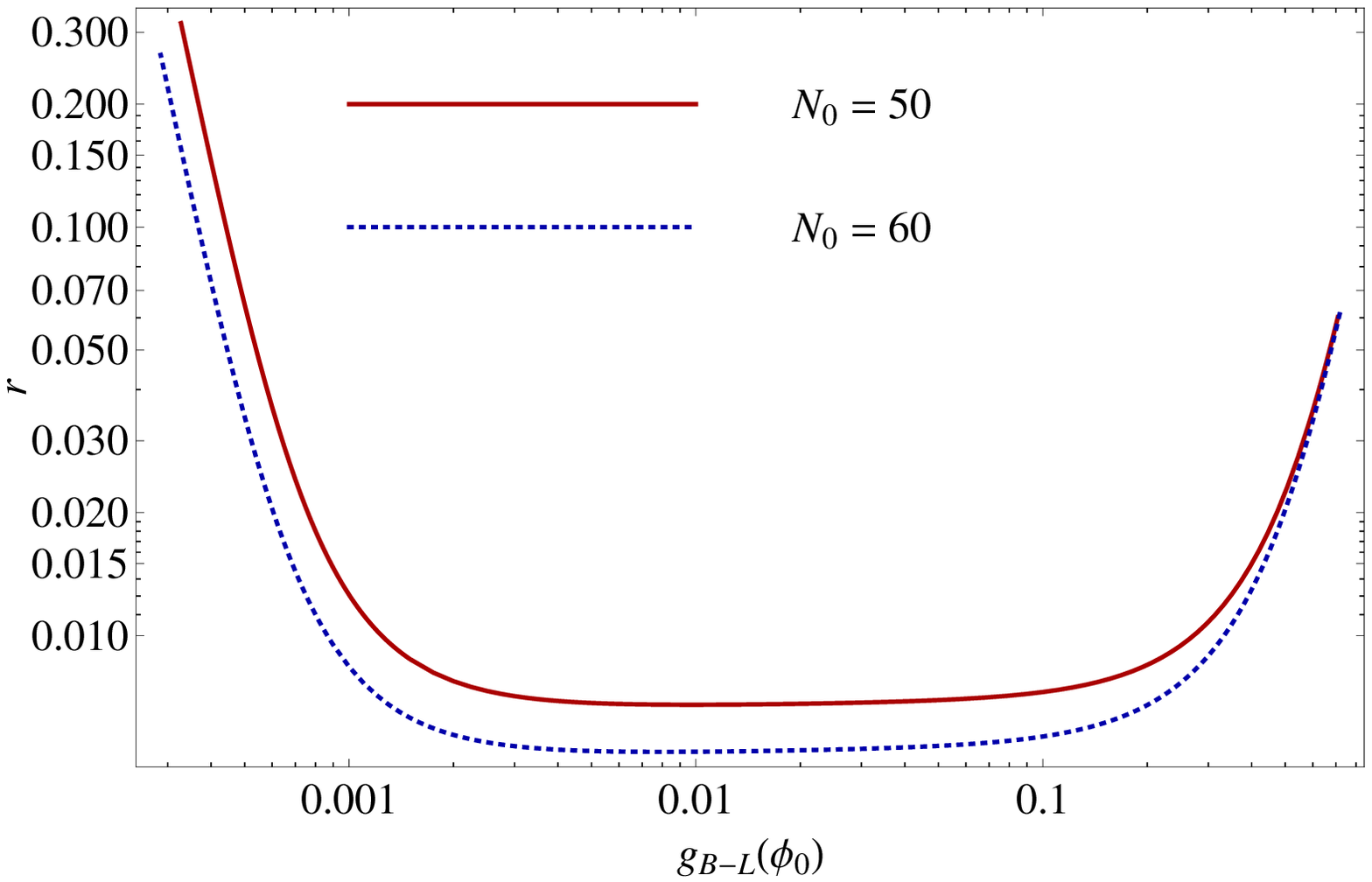}
\centering \includegraphics[width=8.25cm]{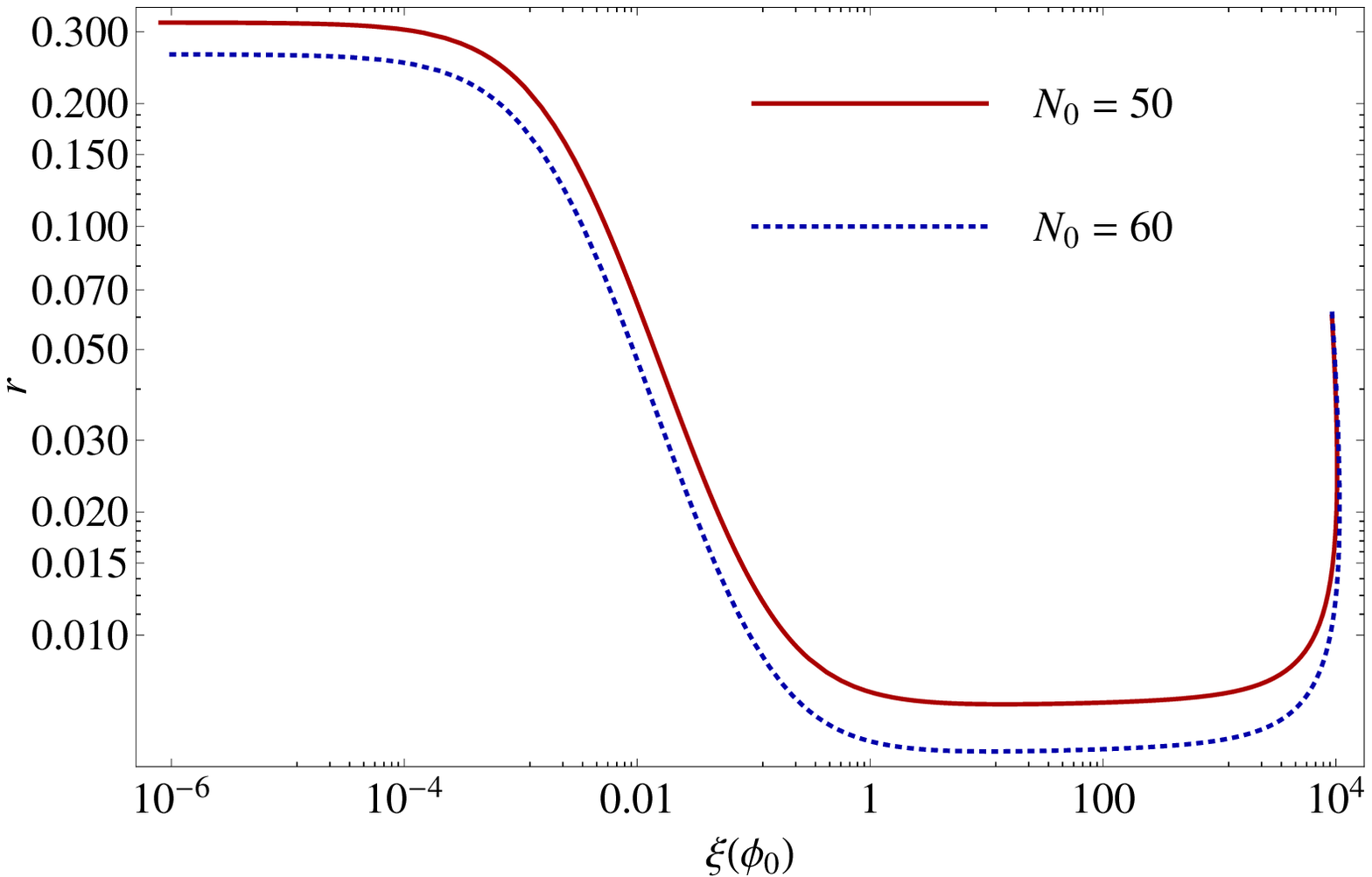}
\caption{$r$ vs. $g_{B-L}(\phi_0)$ and $\xi(\phi_0)$ for 
radiatively corrected non-minimal $\phi^4$ inflation with 
the number of e-foldings $N_0 = 50$ (red solid curve) 
and $N_0 = 60$ (blue dashed curve) 
and $\lambda_0$ = 0.} 
\label{fig3}
\end{figure}

In the large $\xi$ limit, again assuming 
$\lambda_{\phi}$ to be constant, we obtain
the following results for $n_s$, $r$ and 
$\frac{d n_{s}}{d \ln k}$
\bea
n_s &\simeq& 1 - \frac{2}{N_0},  \\
r &\simeq& \frac{12}{N_0^2},  \\
\frac{d n_{s}}{d \ln k} &\simeq&  -\frac{2}{N_0^2},
\eea
with
\beq
\Delta_{\mathcal{R}}^2 \simeq 
\frac{\lambda_{\phi}}{\xi^2}  \left( \frac{N_0^2}{432\,\pi^2}  \right). 
\eeq
We obtain $0.007 \lesssim r \lesssim 0.1$ and
$0.956 \lesssim n_s \lesssim 0.984$ 
consistent with the WMAP 1-$\sigma$ bounds.
The running of the spectral index 
$\frac{d n_{s}}{d \ln k}$ varies from 
$-6 \times 10^{-3}$ to $-8 \times 10^{-3}$. 
Note the second sharp transitions in the predictions 
of $n_s$ and $r$ around $\xi \approx 10^4$. 
Actually, in this limit the approximation 
of the potential given in Eq.~(\ref{ApproxPotential}) 
does not hold as the value of
the gauge coupling $g_{B-L}$ becomes large
and we can no longer ignore its running.

\begin{figure}[t]
\centering \includegraphics[width=8.25cm]{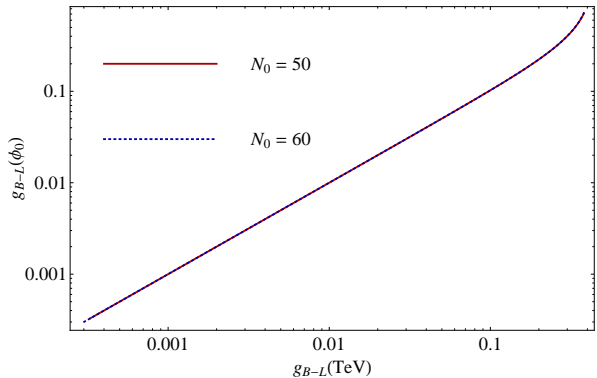}
\caption{$g_{B-L}(\phi_0)$ vs. $g_{B-L}$(TeV) 
for radiatively-corrected non-minimal $\phi^4$ inflation 
with the number of e-foldings $N_0 = 50$ (red solid curve) 
and $N_0 = 60$ (blue dashed curve) and $\lambda_0$ = 0.} \label{fig4}
\end{figure}

\begin{figure}[t]
\centering \includegraphics[width=8.25cm]{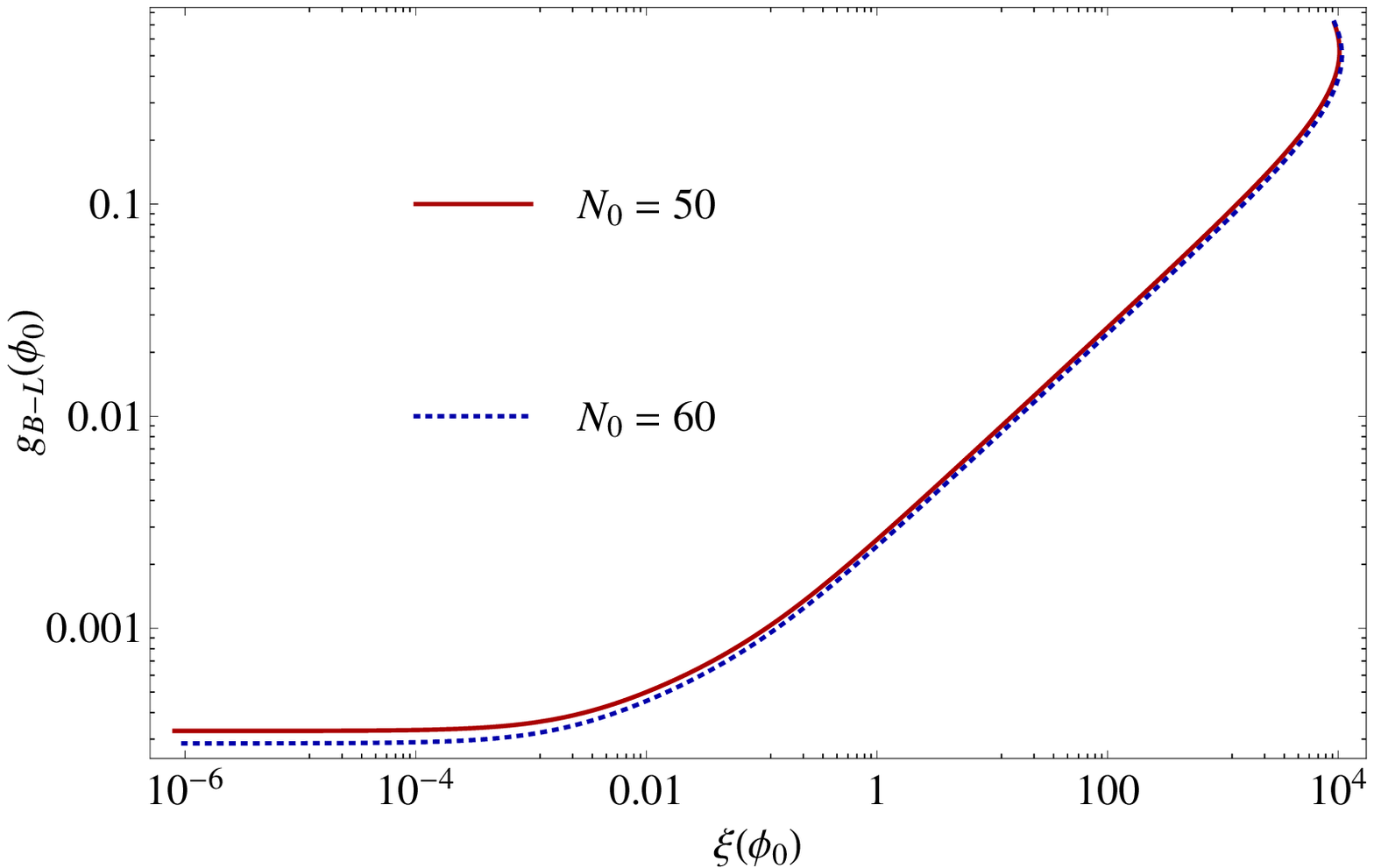}
\caption{$g_{B-L}(\phi_0)$ vs. $\xi(\phi_0)$ 
for radiatively-corrected non-minimal $\phi^4$ inflation 
with the number of e-foldings $N_0 = 50$ (red solid curve) 
and $N_0 = 60$ (blue dashed curve) 
and $\lambda_0$ = 0.} \label{fig5}
\end{figure}

\begin{figure}[h]
\centering \includegraphics[width=8.25cm]{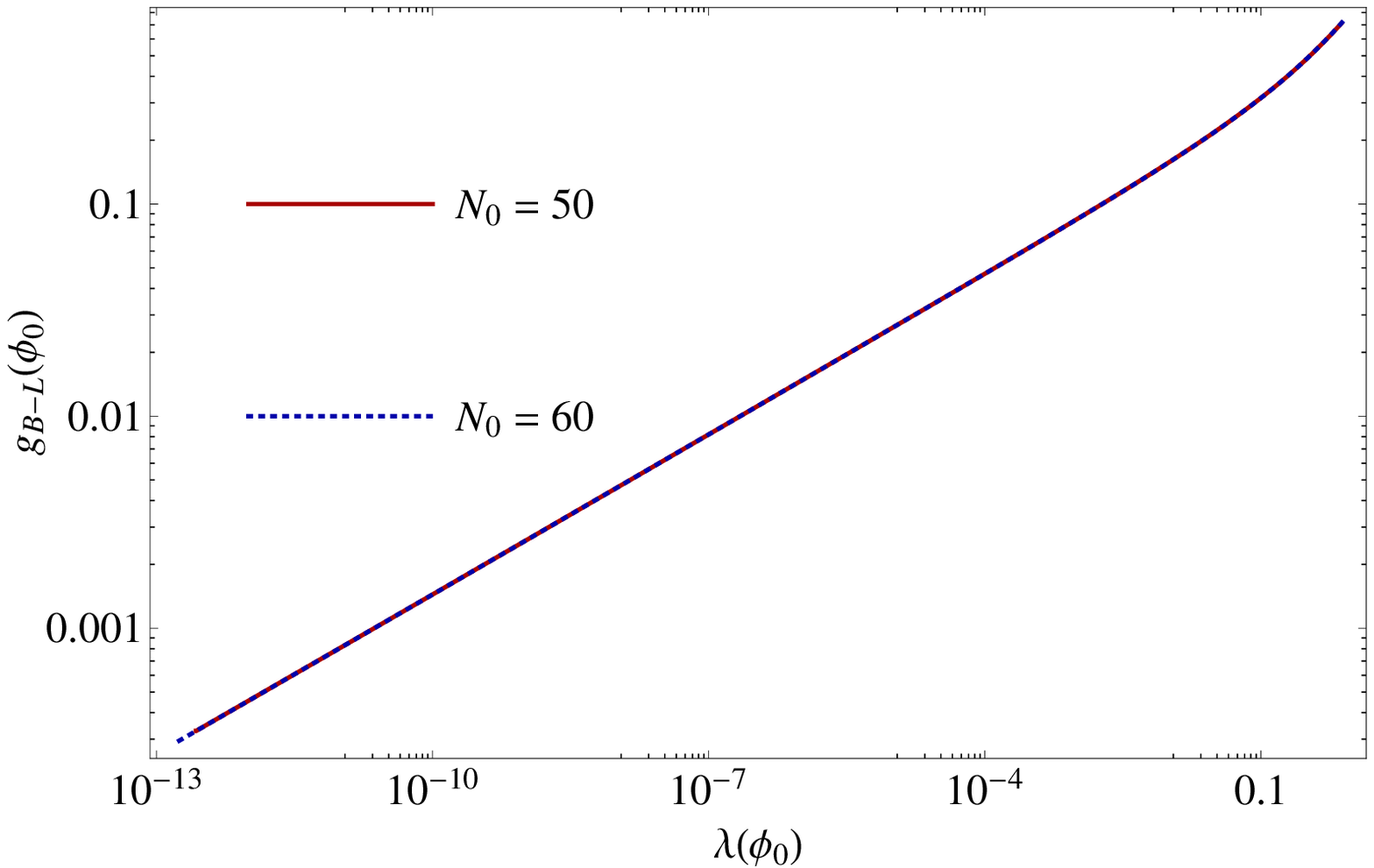}
\caption{$g_{B-L}(\phi_0)$ vs. $\lambda(\phi_0)$ 
for radiatively-corrected non-minimal $\phi^4$ inflation 
with the number of e-foldings $N_0 = 50$ (red solid curve) 
and $N_0 = 60$ (blue dashed curve) 
and $\lambda_0$ = 0.} \label{fig6}
\end{figure}

Finally in Figs.~\ref{fig4}-\ref{fig6} we display 
the relation among the parameters $g_{B-L}(\phi_0)$, 
$g_{B-L}$(TeV),
$\lambda(\phi_0)$ and $\xi(\phi_0)$. Within 1-$\sigma$ 
bounds of WMAP data, these parameters take values 
in the range
$3 \times 10^{-3} \lesssim g_{B-L}(\phi_0) \lesssim 0.46$,
$3 \times 10^{-4} \lesssim g_{B-L}({\rm TeV}) \lesssim 0.32$,
$ 10^{-12} \lesssim \lambda(\phi_0) \lesssim 0.3$ and
$10^{-3}\lesssim \xi(\phi_0) \lesssim 10^4$. 
However, if we require that 
$V^{1/4} \lesssim \Lambda \equiv m_P/\xi$, then more stringent upper bounds,
$g_{B-L}(\phi_0) \lesssim 0.043$, $g_{B-L}({\rm TeV}) \lesssim 0.043$,
$ \lambda(\phi_0) \lesssim 7 \times 10^{-5}$ 
and $\xi(\phi_0) \lesssim 300$ are obtained.
Although, there is some uncertainty in
the calculations of cut-off (i.e., $\Lambda$ 
is argued by some (Ref.~\cite{Bezrukov:2010jz}) to be larger
than $m_P/\xi$ during inflation), 
interesting $B-L$ related LHC physics still looks viable 
even with the more stringent bound on 
$g_{B-L}\,(\lesssim 0.043)$ \cite{Basso:2009hf}.

To summarize, we have considered non-minimal $\lambda\,\phi^4$ 
chaotic inflation in a minimal gauged $U(1)_{B-L}$ extension
of SM. Among the very well-known attractive features 
of this model are the natural presence of three RH neutrinos, 
seesaw mechanism to understand non-zero neutrino masses,
and explanation of the baryon asymmetry via leptogenisis.
With an extra $Z_2$ symmetry one of the three RH neutrino 
can be a viable dark matter candidate.
To realize inflation we utilize the SM gauge singlet inflaton
$\Phi$ which is charged under $B-L$. 
In addition to the non-minimal gravitational coupling, we have 
also included the effect of inflaton-gauge coupling $g_{B-L}$. 
For $ 10^{-12} \lesssim \lambda(\phi_0) \lesssim 0.3$, 
$10^{-3}\lesssim \xi(\phi_0) \lesssim 10^4$ and 
$3 \times 10^{-3} \lesssim g_{B-L}(\phi_0) \lesssim 0.46$
we obtain the inflationary predictions 
$0.956 \lesssim n_s \lesssim 0.984$ and 
$0.007 \lesssim r \lesssim 0.1$
that are consistent with the WMAP 1-$\sigma$ bounds
and will be tested by the Planck satellite.

\begin{figure}[h]
\centering \includegraphics[width=8.25cm]{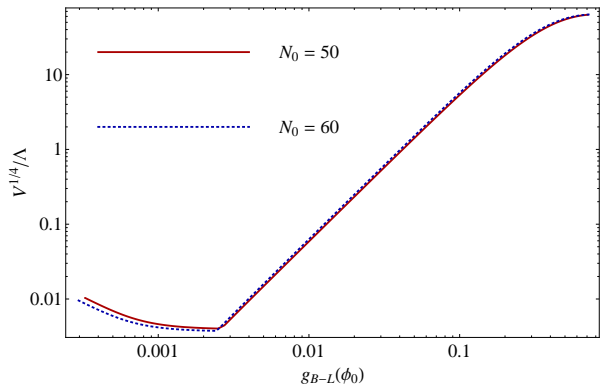}
\caption{$V^{1/4}/\Lambda$ vs. $g_{B-L}(\phi_0)$
for radiatively-corrected non-minimal $\phi^4$ inflation 
with the number of e-foldings $N_0 = 50$ (red solid curve) 
and $N_0 = 60$ (blue dashed curve) 
and $\lambda_0$ = 0.} \label{fig7}
\end{figure}

\section*{Acknowledgments}
%\begin{acknowledgments}
We thank Joshua R. Wickman for valuable discussions.
We also thank Rose Lerner for pointing out the 
importance of including $\xi$ running.
This work is supported in part by the DOE under grant 
No.~DE-FG02-91ER40626 (Q.S. and M.R.) and No.~DE-FG02-10ER41714 (N.O.), 
and by the University of Delaware (M.R.).
N.O. would like to thank the Particle Theory Group of 
the University of Delaware for hospitality during his visit.

%%%%%%%%%%%%%%%%%%%%%%%%%%%%%%%%%

%%%%%%%%%%%%%%%%%%%%%%%%%%%%%%%

\end{document}